\documentclass[twocolumn]{aastex631}

\usepackage{textgreek, gensymb, needspace}
\usepackage[]{fixltx2e}
\usepackage[bottom]{footmisc}

\received{July 6, 2023}
\revised{August 8, 2023}
\accepted{August 18, 2023}

\submitjournal{AJ}

\shorttitle{HST/WFC3 light curve of LTT 1445Ac}
\shortauthors{Pass et al.}

\begin{document}
\title{\small HST/WFC3 Light Curve Supports a Terrestrial Composition for the Closest Exoplanet to Transit an M Dwarf}

\author[0000-0002-1533-9029]{Emily K. Pass}
\affiliation{Center for Astrophysics $\vert$ Harvard \& Smithsonian, 60 Garden Street, Cambridge, MA 02138, USA}

\author[0000-0002-9003-484X]{Jennifer G. Winters}
\affiliation{Center for Astrophysics $\vert$ Harvard \& Smithsonian, 60 Garden Street, Cambridge, MA 02138, USA}
\affiliation{Thompson Physics Lab, Williams College, 880 Main Street, Williamstown, MA 01267, USA}

\author[0000-0002-9003-484X]{David Charbonneau}
\affiliation{Center for Astrophysics $\vert$ Harvard \& Smithsonian, 60 Garden Street, Cambridge, MA 02138, USA}

\author{Aurelia Balkanski}
\affiliation{Center for Astrophysics $\vert$ Harvard \& Smithsonian, 60 Garden Street, Cambridge, MA 02138, USA}

\author[0000-0002-8507-1304]{Nikole Lewis}
\affiliation{Department of Astronomy and Carl Sagan Institute, Cornell University, 122 Sciences Drive, Ithaca, NY 14853, USA}

\author[0000-0002-4443-6725]{Maura Lally}
\affiliation{Department of Astronomy and Carl Sagan Institute, Cornell University, 122 Sciences Drive, Ithaca, NY 14853, USA}

\author[0000-0003-4733-6532]{Jacob L. Bean}
\affiliation{Department of Astronomy \& Astrophysics, University of Chicago, 5640 South Ellis Avenue, Chicago, IL 60637, USA}

\author[0000-0001-5383-9393]{Ryan Cloutier}
\affiliation{Center for Astrophysics $\vert$ Harvard \& Smithsonian, 60 Garden Street, Cambridge, MA 02138, USA}
\affiliation{Department of Physics \& Astronomy, McMaster University, 1280 Main Street West, Hamilton, ON L8S 4L8, Canada}

\author[0000-0003-3773-5142]{Jason D. Eastman}
\affiliation{Center for Astrophysics $\vert$ Harvard \& Smithsonian, 60 Garden Street, Cambridge, MA 02138, USA}



\begin{abstract}
\noindent Previous studies of the exoplanet LTT 1445Ac concluded that the light curve from the Transiting Exoplanet Survey Satellite (TESS) was consistent with both grazing and non-grazing geometries. As a result, the radius and hence density of the planet remained unknown. To resolve this ambiguity, we observed the LTT 1445 system for six spacecraft orbits of the Hubble Space Telescope (HST) using WFC3/UVIS imaging in spatial scan mode, including one partial transit of LTT 1445Ac. This imaging produces resolved light curves of each of the three stars in the LTT 1445 system. We confirm that the planet transits LTT 1445A and that LTT 1445C is the source of the rotational modulation seen in the TESS light curve, and we refine the estimate of the dilution factor for the TESS data. We perform a joint fit to the TESS and HST observations, finding that the transit of LTT 1445Ac is not grazing with 97\% confidence. We measure a planetary radius of 1.07$_{-0.07}^{+0.10}$ R$_\oplus$. Combined with previous radial velocity observations, our analysis yields a planetary mass of $1.37\pm0.19$ M$_\oplus$ and a planetary density of 5.9$_{-1.5}^{+1.8}$ g cm$^{-3}$. LTT 1445Ac is likely an Earth analog with respect to its mass and radius, albeit with a higher instellation, and is therefore an exciting target for future atmospheric studies.
\end{abstract}


\section{Introduction}
\label{sec:intro}
LTT 1445 is a triple star system \citep{Luyten1957, Luyten1980}, located at 6.9 pc and comprising three fully convective M dwarfs. The primary, LTT 1445A, is separated from the close binary LTT 1445BC by 7.2" as of Gaia DR3 \citep{GaiaCollaboration2016, GaiaCollaboration2022}. Photometry from the Transiting Exoplanet Survey Satellite \citep[TESS;][]{Ricker2015} has revealed two small transiting planets in the LTT 1445 system, with the discovery of planet b reported in \citet{Winters2019} and planet c in \citet{Winters2022}. While all three stars fall within the same 21" square TESS pixel, the ground-based transit and radial velocity follow-up presented in those works indicate that the planets orbit the A component. High-precision radial velocity follow-up of LTT 1445A from the ESPRESSO spectrograph \citep{Pepe2021} has also yielded an additional candidate planet d, which is likely non-transiting \citep{Lavie2022}.

LTT 1445A is the nearest M dwarf with a known transiting planet; planets b and c therefore offer some of the most favorable conditions to characterize the atmospheres of terrestrial exoplanets. This favorability can be quantified by the transmission spectroscopy metric \citep[TSM;][]{Kempton2018}, with \citet{Winters2022} calculating a TSM of 30 for planet b and 46 for planet c. \citet{Kempton2018} propose that planets with TSMs substantially larger than 10 are high-priority targets for atmospheric characterization. Moreover, \citet{Winters2019, Winters2022} argue that a future discovery of a planet more favorable for transmission spectroscopy is unlikely based on our understanding of planetary occurrence rates and the fraction of nearby stars already probed by TESS and ground-based surveys. While planet c therefore appears to be an optimal target to devote follow-up resources such as JWST, its TSM was calculated with a caveat: it is possible that the transit of this planet is grazing. A grazing geometry would yield a large uncertainty in the planetary radius, affecting the TSM, our ability to interpret atmospheric observations, and potentially, the terrestrial nature of the planet. While there are non-grazing geometries that are consistent with the TESS data, \citet{Lavie2022} found an 85\% chance that c is grazing. Additional data are needed to resolve this uncertainty and establish whether planet c is suitable for detailed atmospheric characterization.

In this work, we present a transit of LTT 1445Ac as observed by the Hubble Space Telescope (HST) using WFC3/UVIS imaging in spatial scan mode. Unlike the TESS data, these observations resolve LTT 1445A, B, and C as three independent sources. In Section~\ref{sec:data}, we describe the HST data collection and reduction. In Section~\ref{sec:analysis}, we perform a joint fit to the TESS and HST observations. We conclude in Section~\ref{sec:conclusion}.

\section{HST Data Collection and Reduction}
\label{sec:data}

\subsection{Observation setup}

We gathered the HST photometry under program 16503 (PI: Winters), using WFC3/UVIS in spatial scanning mode with the F814W filter and the 512x512 subarray. The WFC3/UVIS spatial scanning mode has not been widely described in the literature, although it has been employed for a handful of projects, including \citet{Riess2014}, \citet{Casertano2016}, \citet{Burke2019}, \citet{Kenworthy2021} and \citet{Fraine2021}.

Our data set comprises six HST orbits, divided between two visits and consisting of sequences of 22s exposures taken at a 76s cadence. The first visit occurred on 2021 Sep 26 from 11:27:24--15:35:39 UT and the second on 2021 Sep 29 from 14:07:57--18:14:44 UT. We observed a partial transit of LTT 1445Ac in the first visit; in the second, the transit fell in the gap between orbital visibility periods.

\begin{figure}[t]
    \centering
    \makebox[\columnwidth][c]{\includegraphics[width=1.\columnwidth]{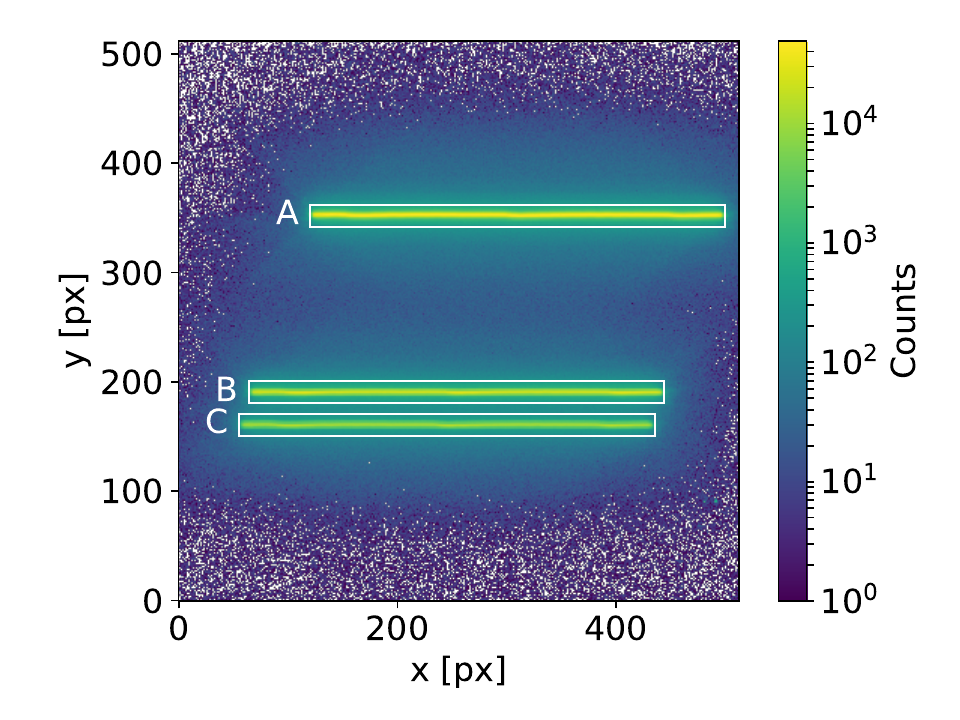}}
    \caption{One of our 188 spatial-scan images of the LTT 1445ABC system. Stars appear as rectangles instead of circles in spatial scans, as the light is smeared in the direction of the scan. This technique allows high-SNR observations of bright sources to be obtained without saturating any pixels. Our 380$\times$20px apertures are indicated in white.}
    \label{fig:apertures}
\end{figure}

\subsection{Preprocessing and extraction}

We download the 188 .flc files associated with our program from the Mikulski Archive for Space Telescopes (MAST). These files are calibrated, individual exposures with charge transfer efficiency corrections applied; for details on the WFC3 calibration pipeline, see appendix E.1 of the WFC3 Instrument Handbook \citep{Dressel_2023}.\footnote{\href{https://hst-docs.stsci.edu/wfc3ihb}{https://hst-docs.stsci.edu/wfc3ihb}}

Next, we use \texttt{WFC3\_phot\_tools} \citep{Shanahan2017},\footnote{\href{https://github.com/cshanahan1/WFC3_phot_tools}{https://github.com/cshanahan1/WFC3\_phot\_tools}} a package designed for WFC3/UVIS spatially scanned data. With this package, we perform cosmic-ray rejection, apply the pixel area map (PAM) correction, and extract each of the three stars using rectangular apertures. We identify appropriate apertures using the \texttt{phot\_tools.detect\_sources\_scan} routine (which uses \texttt{photutils}; \citealt{Bradley2016}) with a SNR threshold of 100. This threshold was chosen to ensure that the algorithm does not combine LTT 1445BC into a single object, which was an issue when using the function's default parameters. We use the x and y centroids determined by this algorithm to center our apertures in each exposure, but use a fixed rectangular aperture size of 380$\times$20 pixels for all three stars (Figure~\ref{fig:apertures}). Our choice of width is motivated by the second visit, in which the spatial scan of A is very close to the edge of the detector; we select the maximal width that does not result in the aperture being truncated by the edge of the detector in any exposure. We explore other choices of aperture size and find that our extraction is robust against modest variations in this choice. We do not perform background subtraction, as the stars are very bright and the background count level is therefore negligible; even towards the edge of the image, the counts are dominated by the tails of the stellar flux, rather than sky background. The HST timestamps are in UTC, which we convert to BJD using \texttt{barycorrpy} \citep{Kanodia2018}.

\begin{figure}[t]
    \centering
    \makebox[\columnwidth][c]{\includegraphics[width=1.2\columnwidth]{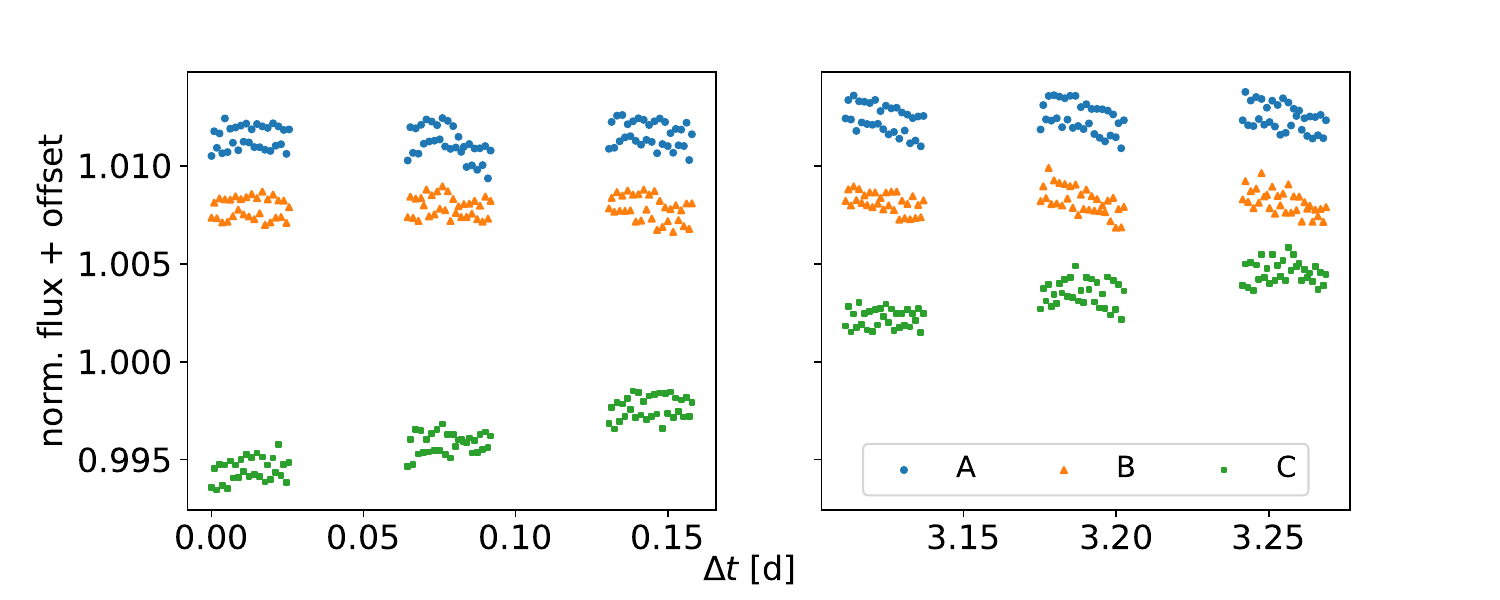}}
    \caption{Our HST aperture photometry prior to removal of systematics. Each star is normalized to its median flux across the two-visit observing campaign. A and B have been shifted by a constant offset such that the light curves do not overlap. The transit occurs at the end of the second orbit.}
    \label{fig:raw}
\end{figure}

Figure~\ref{fig:raw} shows our extracted time series. Some systematics are present. Most prominently, there is an offset between forward and reverse scans. There are also trends with HST orbital phase, with these trends seeming to vary between the first and second visit (the left and right panels of the figure); nonetheless, these trends are much less pronounced than what is typically seen in WFC3/IR spatial scan observations. Figure~\ref{fig:raw} also shows that the flux of LTT 1445C varies substantially, both within a visit and over the 3-day interval between visits. In \citet{Winters2022}, the authors suspected that the C component was the source of the 1.4-day rotation period observed in the TESS data; our observations confirm this hypothesis.

\subsection{Filter comparison}
\label{sec:filter}
\begin{figure}[t]
    \centering
    \makebox[\columnwidth][c]{\includegraphics[width=1.\columnwidth]{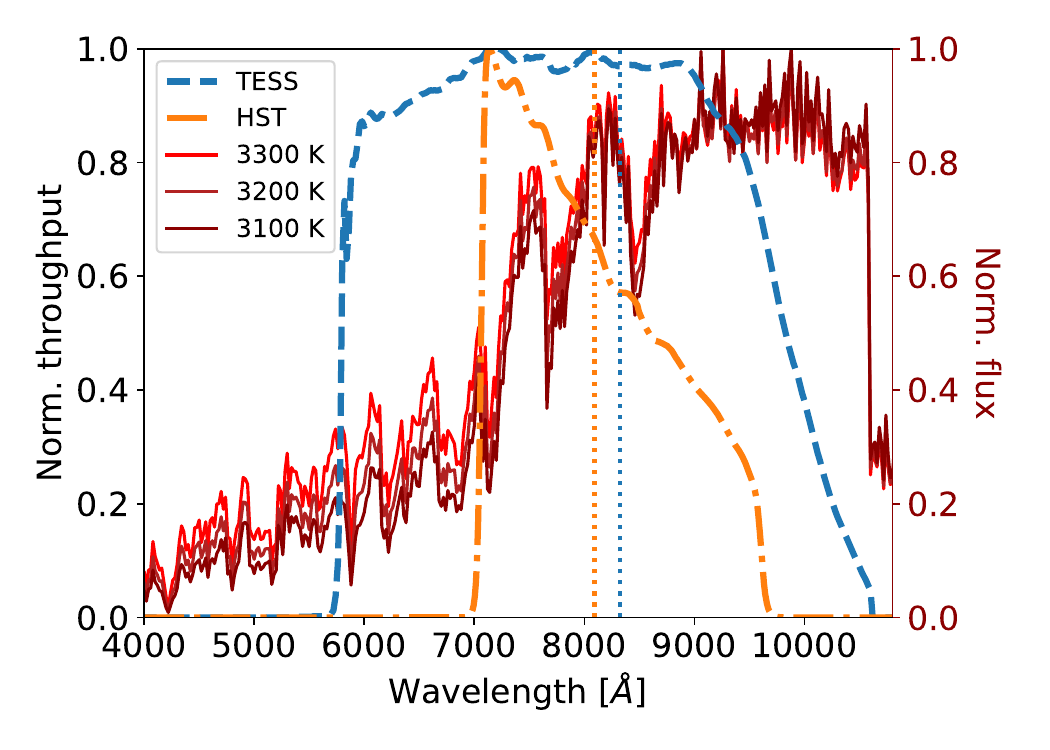}}
    \caption{We compare the TESS response function (blue) to the HST setup used in this work (orange). In red, we overplot the normalized flux for M-dwarf models \citep{Allard2012} for a range of temperatures relevant to the LTT 1445ABC system. For a model 3300K M dwarf (analogous to LTT 1445A), the effective central wavelength of the two instruments is very similar (dotted vertical lines): roughly 8300\AA\ for TESS and 8100\AA\ for HST.}
    \label{fig:band}
\end{figure}

Our HST observations use the F814W filter, which offers good coverage over the red-optical wavelengths where M dwarfs emit much of their light and avoids the 6563\AA\ H\textalpha\ feature, which is sensitive to stellar flares. In Figure~\ref{fig:band}, we compare the response functions of the TESS and HST observations. While the TESS bandpass is wider, the two are nonetheless similar; for a BT-Settl model M dwarf with $T_{\rm eff}$=3300K, log$g$=5, and [Fe/H]=0 \citep{Allard2012}, we find an effective wavelength of 8100\AA\ for our HST observations, as opposed to 8300\AA\ for TESS. The similarity of the response functions allows us to make two simplifications in our analysis, which we describe below.

Firstly, this similarity suggests that we do not need to account for the wavelength dependence of the limb-darkening parameters. To verify this, we use \texttt{exoctk} \citep{Bourque2021} to estimate quadratic limb-darkening parameters for LTT 1445A in each of the two bands, adopting the estimates of the stellar properties from \citet[][]{Winters2019, Winters2022}: $T_{\rm eff}$=3340 K, log$g$=4.967, and [Fe/H]=-0.34 dex. For the TESS bandpass, we find $u_1=0.194, u_2=0.360$. For the HST bandpass, we find $u_1=0.181, u_2=0.367$. The variation in limb darkening between these bandpasses is very small, particularly considering that \citet{Winters2022} allow the limb-darkening parameters to vary in their fit to the TESS data, centered on the computed value and with a standard deviation of 0.1. The parameters are indistinguishable at this level of precision.

A second consideration is the dilution. As the TESS light curve contains all three stars, a dilution correction is required to remove contamination by LTT 1445BC when analyzing those data. In \citet{Winters2019, Winters2022}, the authors estimated this dilution using TESS magnitudes approximated from ($I_{\rm KC}-K_{\rm s}$) colors, ultimately measuring $A_{D} \equiv (f_{\rm A}/( f_{\rm A}+ f_{\rm B} + f_{\rm C}))=0.480\pm0.013$. As HST independently resolves each star, we are equipped to improve this dilution estimate. We measure $A_D= 0.47541 \pm 0.00024$ from our HST observations, with the uncertainty in our measurement driven by spot modulation of LTT 1445C. While not needed for our analysis in this work, we can also measure the dilution factors for the other two stars: 0.3454$\pm$0.0003 for B and 0.1791$\pm$0.0005 for C.

Our HST value for the dilution of A due to BC is consistent with 0.480$\pm$0.013 within errors. Because the TESS and HST bandpasses are relatively similar, we suspect that the differences between them would cause negligible uncertainty in the dilution correction. To test this, we approximate LTT 1445A, B, and C using the 3300K, 3200K, and 3100K models shown in Figure~\ref{fig:band}. We then use these models to calculate $A_D$ in each of the two bands, finding that the estimates differ by only 0.0009. Without any correction for bandpass mismatch, our estimates of $A_D$ from the HST observations still reduce the uncertainty in the TESS dilution by an order of magnitude.

\section{Joint TESS and HST Analysis}
\label{sec:analysis}
\subsection{TESS model}
In \citet{Winters2019, Winters2022}, the authors use the \texttt{exoplanet} package \citep{ForemanMackey2021} to remove stellar variability and \texttt{EXOFASTv2} \citep{Eastman2019} to perform the orbital fit. For simplicity, we perform both tasks with \texttt{exoplanet} in this work. This approach has the benefit of allowing uncertainties in the stellar variability removal to propagate into the final fit.

While we did not observe additional transits of LTT 1445Ab, we fit for both planets in our TESS model. As discussed in \citet{Eastman2022}, transit duration provides an independent measurement of the stellar density that can constrain the stellar radius and mass even beyond the 4--5\% systematic noise floors of current models \citep{Tayar2022}. The inclusion of planet b therefore improves our fit for planet c by reducing the uncertainty in the stellar radius.

We model the rotational modulation with a Gaussian process (GP), using the \texttt{RotationTerm} kernel implemented in \texttt{celerite2} \citep{ForemanMackey2018}. This kernel is a mixture of two simple harmonic oscillators and is parameterized by five hyperparameters: \texttt{sigma}, \texttt{Q0}, \texttt{dQ}, \texttt{f}, and \texttt{period}. As this treatment is fairly standard, we will not discuss the mathematical formalism in greater detail here. We refer the reader to the \texttt{celerite2} documentation for full details, or other works such as \citet{Winters2019}. We adopt the priors on the hyperparameters given in the \texttt{exoplanet} tutorial on this topic:\footnote{\href{https://gallery.exoplanet.codes/tutorials/lc-gp-transit/}{https://gallery.exoplanet.codes/tutorials/lc-gp-transit/}} \texttt{sigma} as an inverse gamma distribution with a lower tail of 1 and upper tail of 5, log(\texttt{Q0}) and log(\texttt{dQ}) as normal distributions with means of 0 and standard deviations of 2, \texttt{f} as a uniform distribution from 0.01 to 1, and log(\texttt{period}) centered at the log of the measured rotation period (here, 1.398 days) with a standard deviation of 0.02. We also fit for a jitter term, which we add to the TESS uncertainties in quadrature. We again adopt the suggested prior from \texttt{exoplanet}: the log of the jitter is a normal distribution centered at the log of the mean TESS error, with a standard deviation of 2.

As in \citet{Winters2019, Winters2022}, we remove the crowding correction applied by the TESS pipeline and apply our own. While those works used $A_D = 0.480$, we instead use $A_D = 0.47541$, as measured from our HST observations in Section~{\ref{sec:filter}}.

We use \texttt{PyMC3} \citep{Salvatier2016} as our modeling framework. As in \citet{Winters2019, Winters2022}, we use the \citet{Benedict2016} $K$-band mass--luminosity relation as our prior on the stellar mass, which produces an estimate of 0.258$\pm$0.014 M$_\odot$, and the \citet{Boyajian2012} mass--radius relation as our prior on the stellar radius, corresponding to 0.268$\pm$0.027 R$_\odot$. We model the star as an \texttt{exoplanet.LimbDarkLightCurve} object, which accounts for quadratic limb darkening. We use the limb-darkening estimates from Section~\ref{sec:filter} as our priors: normal distributions centered at $u_1 = 0.19$ and $u_2 = 0.36$ and with standard deviations of 0.10.

We model the light curve corresponding to the two planets using the \texttt{exoplanet.orbits.KeplerianOrbit} class, which takes as input the planetary periods, times of conjunction, impact parameters, radius ratios, and transit durations, as well as the stellar radius. We also include a free parameter for the normalization of the light curve. This class outputs an estimate of stellar mass implied by the transit duration of each planet; we use these estimates as \texttt{PyMC3} observed parameters, comparing them against our mass prior.

As \citet{Winters2022} and \citet{Lavie2022} both found that the orbits of planets b and c were consistent with circular, we do not include eccentricity in our fit. Moreover, low eccentricities are expected due to tidal circularization \citep[e.g.,][]{Adams2006, Jackson2008}: for Earth-like tidal dissipation factors, the circularization timescale for both planets is on the order of 1--10 million years. While ages are challenging to measure for M dwarfs, the long rotation period and H\textalpha\ inactivity of LTT 1445A rule out such extreme youth \citep[e.g.,][]{Medina2022, Pass2022}.

\begin{figure*}[t]
    \centering
    \makebox[\textwidth][c]{\includegraphics[width=0.85\textwidth]{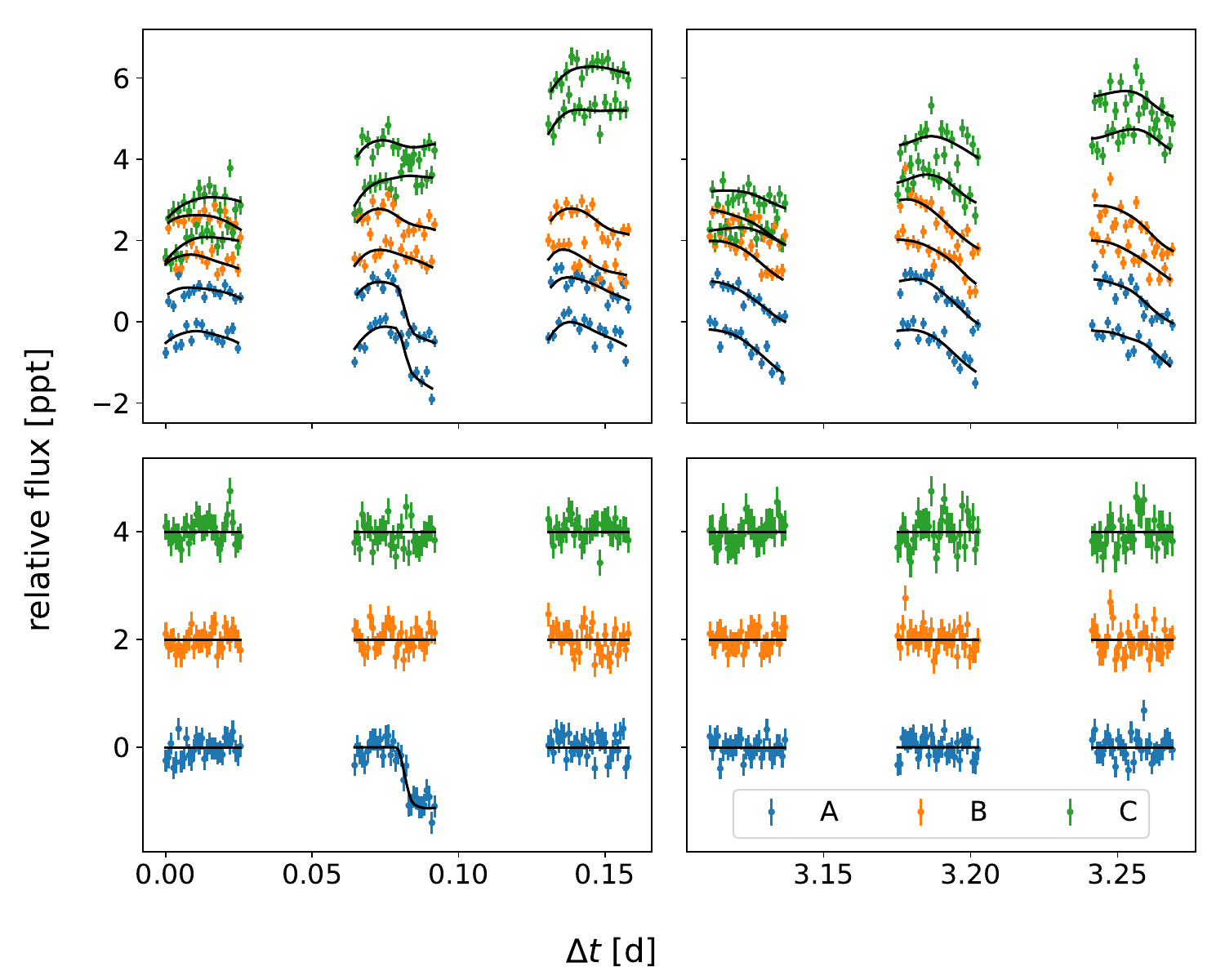}}
    \caption{Our MAP fit to the HST observations. In the upper panels, we show the joint systematics and transit fit to the raw data. In the lower panels, we apply the systematics correction; the resulting light curves contain only residual noise and, in the case of A, the transit. The error bars in the upper panel show photon noise, while in the lower panel we also include our estimated jitter term in quadrature. A y-axis offset has been applied to each light curve for clarity.}
    \label{fig:syst}
\end{figure*}

We use uninformative uniform priors for the radius ratios and transit durations. For the impact parameter, we use a uniform prior ranging between 0 and $1+R_P/R_*$; an impact parameter exceeding this upper limit would correspond to a completely non-transiting planet. For the periods and times of conjunction, we use uniform priors centered at the \citet{Winters2022} value with a width of 0.02 days; this width is chosen to be substantially larger than the uncertainties estimated in \citet{Winters2022}, ensuring that our solutions can deviate from those results if necessitated by the data.

\subsection{HST model}
We fit our HST and TESS data jointly, and therefore the planetary and stellar parameters described in the previous section are also used to model the HST transit. As discussed in Section~\ref{sec:filter}, we do not find it necessary to use separate limb-darkening parameters for the HST observations, as we find that the differences between the bandpasses will not measurably change the limb darkening at the level of precision of our data. We perform our systematics correction within the \texttt{PyMC3} fit to allow uncertainties in the correction to propagate into our inferred system parameters.

\begin{deluxetable}{lccl}[t]
\tabletypesize{\scriptsize}
\tablecaption{Raw and Systematics-corrected Light Curves of the LTT 1445 system\label{tab:flux}}
\tablehead{ 
\colhead{Column} & 
\colhead{Format} &  
\colhead{Units} & 
\colhead{Description}} 
\startdata 
1 & F9.5 & days & BJD - 2457000\\
2 & F7.0 & counts & Raw Flux A \\
3 & F7.0 & counts & Raw Flux B \\ 
4 & F7.0 & counts & Raw Flux C \\ 
5 & F3.3 & ppt & Corrected Flux A \\
6 & F3.3 & ppt & Error in Corrected Flux A \\ 
7 & F3.3 & ppt & Corrected Flux B\\
8 & F3.3 & ppt & Error in Corrected Flux B \\
9 & F3.3 & ppt & Corrected Flux C \\ 
10 & F3.3 & ppt & Error in Corrected Flux C\\\enddata
\tablecomments{Full table available in machine-readable form. The corrected flux columns use the maximum \textit{a posteriori} systematics correction, as plotted in Figure~\ref{fig:syst}.\newline \newline}
\end{deluxetable}

\begin{deluxetable}{lc}[t]
\tabletypesize{\footnotesize}
\tablecolumns{2}
\tablewidth{100pt}
\tablecaption{Maximum \textit{a posteriori} Jitter Parameters \label{tab:jitter}}
\tablehead{
\colhead{Jitter Parameter} & 
\colhead{ppt}}
\startdata
\hphantom{00000000000000}A, HST Visit 1\hphantom{00000000000000} & 0.15 \\ 
\hphantom{00000000000000}B, HST Visit 1\hphantom{00000000000000} & 0.15 \\ 
\hphantom{00000000000000}C, HST Visit 1\hphantom{00000000000000} & 0.09 \\
\hphantom{00000000000000}A, HST Visit 2\hphantom{00000000000000} & 0.15 \\
\hphantom{00000000000000}B, HST Visit 2\hphantom{00000000000000} & 0.17 \\
\hphantom{00000000000000}C, HST Visit 2\hphantom{00000000000000} & 0.17 \\
\hphantom{00000000000000}TESS\hphantom{00000000000000} & 0.62 \\
\enddata
\tablecomments{These jitter terms are added to the observational uncertainties in quadrature.\newline}
\end{deluxetable}
\vspace{-0.85cm}

\begin{figure}[t]
    \centering
    \makebox[\columnwidth][c]{\includegraphics[width=1.2\columnwidth]{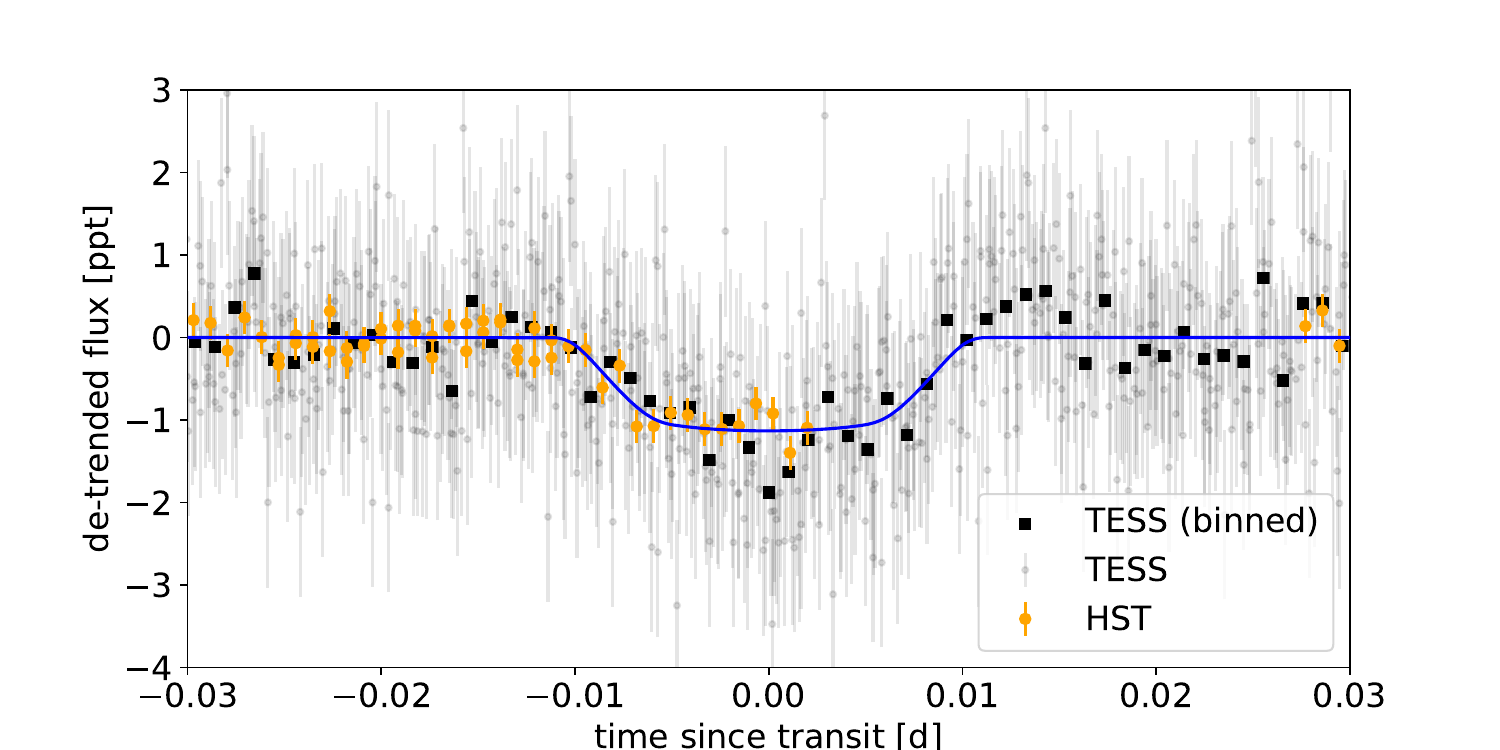}}
    \caption{The phased light curve of LTT 1445A, highlighting the transit of planet c. The detrended TESS observations are show in black and our new HST observations in orange. In blue, we show our MAP orbital solution.}
    \label{fig:transit}
\end{figure}

For each star, we compare the normalized observed fluxes to our model. This model comprises the \texttt{exoplanet.orbits.KeplerianOrbit} prediction (which is always equal to unity for stars B and C), divided by a normalized systematics term. This systematics term is a fourth-order polynominal in HST orbital phase, with coefficients that are shared between the three stars but allowed to vary between the two visits. The normalization of the observed fluxes is also a free parameter, which is allowed to vary between stars, visits, and forward/reverse scans. For LTT 1445C, this normalization includes a linear slope to model the observed rotational modulation, which we allow to vary between visits. Lastly, we include a GP to clean up residual correlations that we observe between the light curves of the three stars. The kernel for this GP is a \texttt{celerite2} stochastically driven, damped harmonic oscillator, with two hyperparameters that govern the characteristic amplitude and length scale of oscillations. These hyperparameters are shared across all three stars and across both visits. We also fit for a jitter term that is added to our photometric uncertainties in quadrature. This jitter is allowed to vary between stars and between visits.

The maximum \textit{a posteriori} (MAP) solution for our systematics correction is shown in Figure~\ref{fig:syst}, with the corresponding flux values provided in Table~\ref{tab:flux}. In Figure~\ref{fig:transit}, we show the phased transit of LTT 1445Ac as it appears in the detrended TESS and HST data. In Table~\ref{tab:jitter}, we note the MAP values for the jitter terms included in our model. The standard deviation of the residuals of our MAP fit is 188 ppm for the HST observations of A (22s exposures) and 1.02 ppt for the TESS observations (2m exposures).

Inspection of the residuals in Figure~\ref{fig:syst} reveals that there is some correlated noise in the systematics-corrected HST light curves, likely resulting from intrinsic stellar variation. The jitter terms included in our model have increased the error bars of each observation to account for this noise. However, this treatment may be problematic: absorbing correlated noise into white-noise jitter can sometimes bias parameter estimates \citep[e.g.,][]{Aigrain2022}. We therefore also consider a version of our model that does not include jitter terms, instead modelling the correlated noise for each star using separate GPs (again, the kernels are \texttt{celerite2} stochastically driven, damped harmonic oscillators with two hyperparameters). The parameters we state in the remainder of this manuscript are determined using this red-noise model. However, the distinction is ultimately unimportant: after performing the sampling described in the following section, we find that the two models produce consistent estimates of the system parameters and their uncertainties.

\begin{deluxetable*}{lccccccc}
\tablecaption{Median Parameters and 68\% Confidence Intervals from our Joint HST--TESS Fit}
\tablecolumns{4}
\tablehead{\colhead{Parameter} & \colhead{Description} &  \multicolumn{2}{c}{Values} }
\startdata
\multicolumn{2}{l}{Host star parameters:}&LTT~1445A\smallskip\\
~~~~$M_{*}$\dotfill &Stellar mass (M$_\odot$) \dotfill& $0.257\pm0.014$\\
~~~~$R_{*}$\dotfill &Stellar radius (R$_{\odot}$) \dotfill&  $0.271^{+0.019}_{-0.010}$ \\
~~~~$u_{1}$\dotfill &Linear limb-darkening coefficient \dotfill & $0.158^{+0.082}_{-0.077}$\\
~~~~$u_{2}$\dotfill &Quadratic limb-darkening coefficient \dotfill & $0.34^{+0.10}_{-0.09}$\\
\hline
\multicolumn{2}{l}{Planetary parameters:}&c&b\smallskip\\
~~~~$P$\dotfill &Period (days)\dotfill & $3.1238994^{+0.0000031}_{-0.0000033}$&$5.3587635^{+0.0000044}_{-0.0000045}$\\
~~~~$a$\dotfill &Semimajor axis (au)\dotfill & $0.02659^{+0.00047}_{-0.00049}$ & $0.03810^{+0.00067}_{-0.00070}$\\
~~~~$T_0$\dotfill &Time of conjunction (BJD)\dotfill & $2458412.58218^{+0.00078}_{-0.00074}$&$2458412.70910^{+0.00047}_{-0.00046}$ \\  
~~~~$T_{14}$\dotfill &Total transit duration (days)\dotfill & $0.02101^{+0.00091}_{-0.00088}$&$0.05691^{+0.00080}_{-0.00075}$ \\
~~~~$R_P$\dotfill &Radius (R$_\oplus$)\dotfill & $1.07^{+0.10}_{-0.07}$&$1.34^{+0.11}_{-0.06}$\\
~~~~$R_P/R_*$\dotfill &Radius of planet in stellar radii \dotfill & $0.0362^{+0.0019}_{-0.0016}$&$0.0454\pm0.0012$ \\
~~~~$\delta$\dotfill &Transit depth (fraction)\dotfill & $0.001104^{+0.000076}_{-0.000072}$&$0.00230\pm{0.00011}$ \\
~~~~$i$\dotfill &Inclination ($\degree$)\dotfill & $87.46^{+0.13}_{-0.21}$&$89.53^{+0.33}_{-0.40}$ \\
~~~~$b$\dotfill &Impact parameter \dotfill & $0.937^{+0.012}_{-0.011}$&$0.25^{+0.18}_{-0.17}$ \\
~~~~$b+R_P/R_*$\dotfill &Grazing parameter\dotfill & $0.973^{+0.013}_{-0.011}$&$0.30^{+0.18}_{-0.17}$ \\
~~~~$K$\dotfill &RV semi-amplitude (ms$^{-1}$)\dotfill & $1.48\pm0.20$ & $2.47\pm0.20$ \\
~~~~$M_P$\dotfill &Mass (M$_\oplus$)\dotfill & $1.37\pm0.19$ & $2.73^{+0.25}_{-0.23}$ \\
~~~~$\rho_P$\dotfill &Density (g cm$^{-3}$)\dotfill & $5.9^{+1.8}_{-1.5}$&$6.2^{+1.2}_{-1.3}$\\
\hline
\multicolumn{2}{l}{Derived parameters, assuming $T_{\rm eff}=3340\pm150$~K \citep{Winters2022}:}\smallskip\\
~~~~$T_{\rm eq}$\dotfill &Zero-albedo equilibrium temperature (K)\dotfill & $516^{+28}_{-27}$&$431\pm23$\\
~~~~$S$\dotfill &Instellation ($S_\oplus$)\dotfill & $11.7^{+2.7}_{-2.3}$&$5.7^{+1.3}_{-1.1}$\\
~~~~TSM\dotfill &Transmission spectroscopy metric\dotfill & $42^{+12}_{-9}$&$34.5^{+6.4}_{-4.9}$\\
\enddata
\vspace{0.6cm}
\label{tab:master_table}
\end{deluxetable*}
\vspace{-0.92cm}

\subsection{Sampling}
We run a Markov-Chain Monte Carlo (MCMC) to determine the uncertainties in our model parameters. Starting from the MAP solution found by \texttt{exoplanet}, we use the modified \texttt{PyMC3} sampler implemented in \texttt{exoplanet} to sample four chains each with a 1500-draw burn-in and 2000 draws. We use an initial acceptance fraction of 0.5, a target acceptance fraction of 0.95, and 100 regularization steps. We find that the sampler properly converges, as evidenced by Gelman–Rubin statistics \citep{Gelman1992} near 1 for all parameters (specifically, all parameters are within a tolerance of 0.003).

\begin{figure}[t]
    \centering
    \makebox[\columnwidth][c]{\includegraphics[width=\columnwidth]{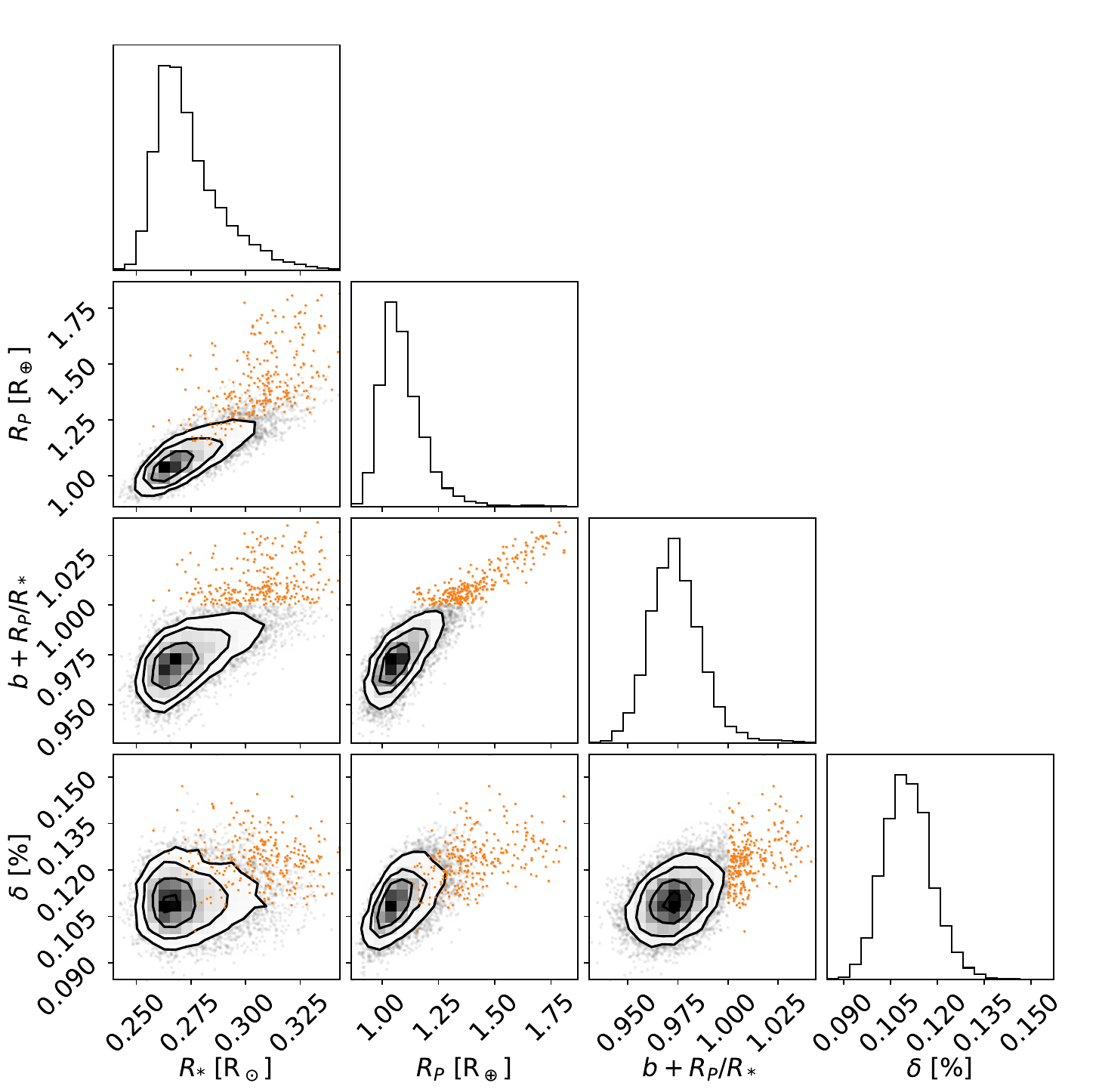}}
    \caption{A corner plot showing our posterior distributions for some key parameters of our LTT 1445Ac fit. In orange, we highlight the 3\% of samples consistent with a grazing geometry. The grazing solutions tend towards large impact parameters, transit depths, and stellar radii.}
    \label{fig:corner}
\end{figure}

In Table~\ref{tab:master_table}, we tabulate our median and 68\% confidence intervals for the orbital parameters. We are particularly interested in the grazing parameter, $b$+$R_P/R_*$. When this parameter exceeds 1, only part of the planet eclipses the stellar disk during transit and the transit is considered grazing. We measure a value of $0.973^{+0.013}_{-0.011}$, indicating a non-grazing geometry. Moreover, only 3\% of samples in our posterior distribution have a grazing parameter in excess of 1 (Figure~\ref{fig:corner}). We are therefore able to constrain the radius of LTT 1445Ac with good precision, finding a value of $1.07^{+0.10}_{-0.07}$ R$_\oplus$. This measurement is consistent with the $1.147^{+0.055}_{-0.054}$ R$_\oplus$ reported in \citet{Winters2022}, although that value was estimated using the \citet{Chen2017} planetary mass--radius relation as a prior and not measured solely from the light curve, explaining its smaller uncertainty.

\subsection{Radial velocities}
For circular orbits, the correlation between the transit parameters and the RV semiamplitude, $K$, is negligible. As we have not collected any new RV data, we can adopt the measurement of $K$ from a previous work. However, no previous work has performed a radial-velocity fit using all extant RV data. \citet{Winters2022} analyzed 136 radial velocities from the ESPRESSO \citep{Pepe2021}, HARPS \citep{Mayor2003}, HIRES \citep{Vogt1994}, MAROON-X \citep{Seifahrt2016, Seifahrt2018, Seifahrt2020}, and PFS \citep{Crane2006, Crane2008, Crane2010} spectrographs, finding $K_c=1.67^{+0.21}_{-0.20}$~ms$^{-1}$. \citet{Lavie2022} collected 85 additional radial velocities with ESPRESSO and found $K_c=1.11$$\pm$$0.20$ ms$^{-1}$, but they analyzed only their ESPRESSO data and archival data from HARPS. The two estimates do not agree within stated errors. While the \citet{Lavie2022} fit also includes a third planet, they report that their measurement of $K_c$ is effectively unchanged in a model fit that contains only planets b and c; therefore, the inclusion or exclusion of planet d does not explain the discrepancy.

Repeating the \texttt{EXOFASTv2} RV-only analysis from \citet{Winters2022} but with the addition of the 85 new ESPRESSO radial velocities from \citet{Lavie2022}, we find $K_c=1.48$$\pm$$0.20$ ms$^{-1}$, which splits the difference between the previous measurements. This fit assumes circular orbits and uses separate zero points for the observations collected before and after the COVID-19 shutdown of ESPRESSO. For planet b, this analysis produces $K_b=2.47$$\pm$$0.20$ ms$^{-1}$, which also is intermediate between the 2.60$\pm$0.21 ms$^{-1}$ found by \citet{Winters2022} and the 2.15$\pm$0.19 ms$^{-1}$ found by \citet{Lavie2022}. We prefer these new estimates over those published in the previous works, as they account for all extant radial velocity data. We combine these $K$ measurements with our posterior distributions of period, inclination, and stellar mass to estimate the planetary masses, finding $1.37\pm0.19$~M$_\oplus$ for planet c and $2.73^{+0.25}_{-0.23}$~M$_\oplus$ for planet b.

\subsection{Discussion}
\begin{figure}[t]
    \centering
    \makebox[\columnwidth][c]{\includegraphics[width=1.1\columnwidth]{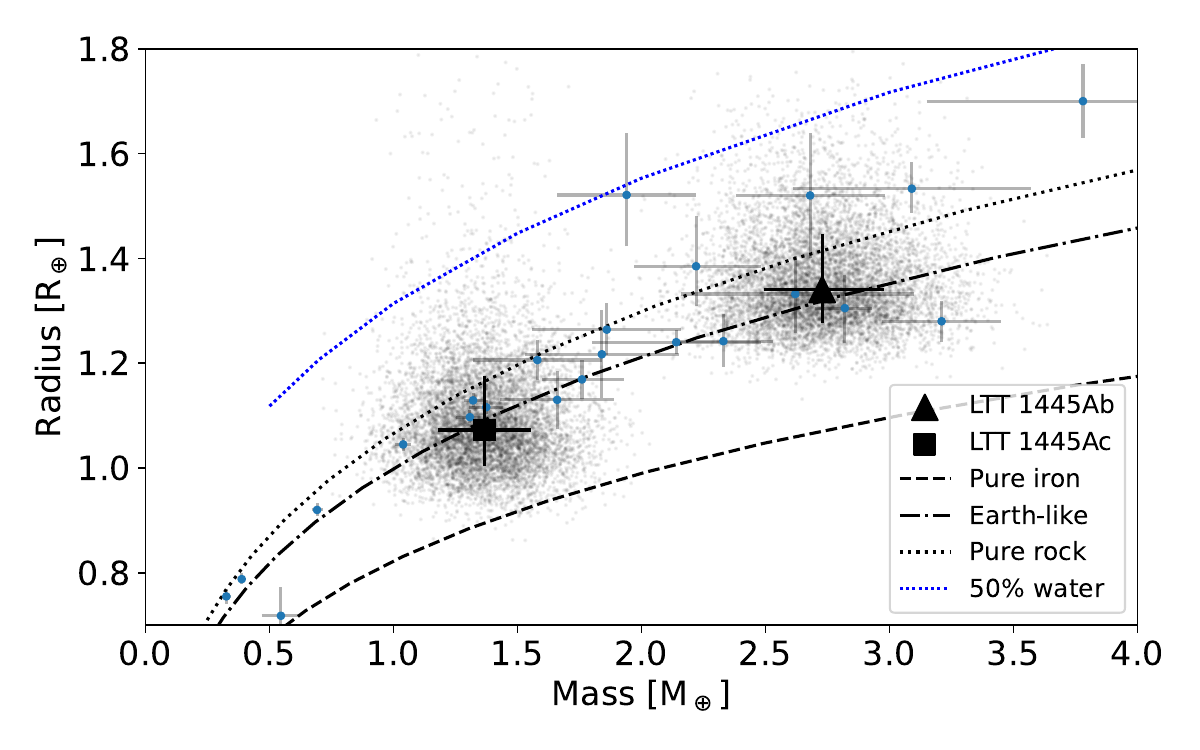}}
    \caption{Our mass and radius estimates for LTT 1445Ab and c, including our posterior samples. We overplot composition curves from \citet{Zeng2019}, illustrating the mass--radius relation for 100\% rock, 100\% iron, Earth-like (32.5\% Fe+67.5\% MgSiO$_3$), and water-world (50\% water+50\% Earth-like) compositions. Both planets fall along the Earth-like composition curve. For comparison, we also include other small planets around M dwarfs ($<0.6$M$_\odot$) with planetary masses and radii measured to better than 20\% error, as tabulated in the NASA Exoplanet Archive \citep[][accessed 2023 May 22]{Akeson2013}.}
    \label{fig:mr}
\end{figure}

The densities implied by our mass and radius estimates are fully consistent with Earth-like planetary compositions (Figure~\ref{fig:mr}). We find that 78.6\% of our posterior samples fall below the pure-rock line for LTT 1445Ac and 75.6\% for LTT 1445Ab, indicating that a minority of our solutions favor the inclusion of some water or H/He. \citet{Luque2022} identified a population of M-dwarf “water-world” exoplanets that they suggest are composed of a 1:1 ratio of water to rock. It is possible but unlikely that the LTT 1445A planets are members of this population: we find that 98.2\% of our planet c samples fall below the 50\%-water composition curve, and 99.1\% for planet b.

Both planets are more highly irradiated than the Earth, with an instellation of roughly 12$S_\oplus$ for planet c and 6$S_\oplus$ for planet b.

\section{Conclusion}
\label{sec:conclusion}

We observed the three stars of the LTT 1445 system for six orbits of the Hubble Space Telescope using WFC3/UVIS imaging in spatial scan mode, including one transit of LTT 1445Ac. We jointly fit our observations with extant TESS data, allowing us to establish that the transit of LTT 1445Ac is non-grazing with 97\% confidence and measure the planetary radius to be 1.07$_{-0.07}^{+0.10}$ R$_\oplus$. Using radial velocity observations previously published in \citet{Winters2022} and \citet{Lavie2022}, we find a planetary mass of $1.37\pm0.19$~M$_\oplus$.

We tabulate our constraints on planetary parameters for LTT 1445Ab and c in Table~\ref{tab:master_table}. These estimates supersede those of \citet{Winters2022}, as ours include additional transit data from HST and radial velocities from \citet{Lavie2022} and are not calculated under the assumption of the \citet{Chen2017} mass--radius relation. Our revised system parameters yield a TSM of 42 for LTT 1445Ac, similar to the value estimated in \citet{Winters2022} using that mass--radius assumption. Our HST observations also allow us to confirm that LTT 1445C is the source of the rotational modulation in the TESS observations and refine the estimate of the TESS dilution to $A_D=0.4754$.

Taken together, our inferred mass and radius indicate that LTT 1445Ac has a likely terrestrial composition, falling on the rocky side of the radius gap \citep{Fulton2017}. As the nearest terrestrial exoplanet to transit an M dwarf (alongside LTT 1445 Ab), this planet is an exciting target for atmospheric characterization, particularly now that it is known to be non-grazing and its radius is therefore appropriately constrained.

\section*{Acknowledgements}
We thank Jonathan Irwin and Johanna Teske for their feedback on both the HST proposal and this manuscript, Nicola Astudillo-Defru, Xavier Bonfils, Martti Holst Kristiansen, Andrew Howard, Alton Spencer, and Andrew Vanderburg for their participation in the HST proposal, and the anonymous referee for their helpful comments that improved this paper. E.P.\ is supported in part by a Natural Sciences and Engineering Research Council of Canada (NSERC) Postgraduate Scholarship, M.L.\ by a National Science Foundation (NSF) Graduate Research Fellowship, and R.C.\ by an NSERC Banting Postdoctoral Fellowship.

This work is based on observations with the NASA/ESA Hubble Space Telescope obtained at the Space Telescope Science Institute, which is operated by the Association of Universities for Research in Astronomy, Incorporated, under NASA contract NAS5-26555. Support for program number HST-GO-16503 was provided through a grant from the STScI under NASA contract NAS5-26555. This paper includes data collected by the TESS mission, which are publicly available from the Mikulski Archive for Space Telescopes (MAST). Funding for the TESS mission is provided by the NASA's Science Mission Directorate.

\facilities{HST, TESS}

\software{\texttt{barycorrpy} \citep{Kanodia2018}, \texttt{corner} \citep{ForemanMackey2016}, \texttt{exoctk} \citep{Bourque2021}, \texttt{matplotlib} \citep{Hunter2007}, \texttt{numpy} \citep{Harris2020}, \texttt{stsynphot} \citep{SDT2020}, \texttt{WFC3\_phot\_tools} \citep{Shanahan2017}, as well as \textsf{exoplanet} \citep{ForemanMackey2021,
ForemanMackey2021a} and its dependencies \citep{ForemanMackey2017,
ForemanMackey2018, Agol2020, Kumar2019,
AstropyCollaboration2013, AstropyCollaboration2018, Kipping2013,
Luger2019, Salvatier2016, TDT2016}.}

\bibliography{hst}{}
\bibliographystyle{aa_url}



\end{document}